# Transforming Higher Education with AI-Powered Video Lectures

Dengsheng Zhang

Institute of Innovation, Science and Sustainability

Federation University Australia

dengsheng.zhang@federation.edu.au

## Abstract:

The integration of artificial intelligence (AI) into video lecture production has the potential to transform higher education by streamlining content creation and enhancing accessibility. This paper investigates a semi-automated workflow that combines Google Gemini for script generation, Amazon Polly for voice synthesis, and Microsoft PowerPoint for video assembly. Unlike fully automated text-to-video platforms, this hybrid approach preserves pedagogical intent while ensuring script–slide synchronization, narrative coherence, and customization. Case studies demonstrate the effectiveness of Gemini in generating accurate and context-sensitive scripts for visually rich academic presentations, while Polly provides natural-sounding narration with controllable pacing. A two-course pilot study was conducted to evaluate AI-generated instructional videos (AIIV) against human instructional videos (HIV). Both qualitative and quantitative results indicate that AIIVs are comparable to HIVs in terms of learning outcomes, with students reporting high levels of clarity, coherence, and usability. However, limitations remain, particularly regarding audio quality and the absence of human-like avatars. The findings suggest that AI-assisted video production can reduce instructor workload, improve scalability, and deliver effective learning resources, while future improvements in synthetic voices and avatars may further enhance learner engagement.

# 1. Introduction

The integration of artificial intelligence (AI) into educational technologies has revolutionized the production and delivery of instructional materials. Among these innovations, AI-assisted video lecture creation stands out as a promising avenue for enhancing accessibility, personalization, and scalability in teaching. Traditional methods of recording video lectures are often time-consuming and resource-intensive, requiring multiple takes, post-production editing, and substantial instructor effort. By contrast, AI-based tools now offer streamlined workflows where scripts can be generated via large language models like Google Gemini, voices synthesized with tools such as Amazon Polly, and visuals seamlessly integrated into slide-based video formats.

Recent research provides a broad perspective on the potential and limitations of AI-generated instructional videos. Netland et al. (2025) conducted a comparative study between human-made and AI-generated teaching videos, concluding that while learning outcomes were comparable, learners perceived human videos as more engaging and personal. Similarly, Xu et al. (2024) found that students performed equally well with AI-generated videos but reported mixed experiences regarding emotional connection and delivery quality. These studies suggest that while AI videos are effective in terms of information transfer, emotional resonance remains a challenge.

Other studies such as Leiker et al. (2023), Pellas (2025), Arkün-Kocadere and Özhan (2024), delve deeper into the pedagogical implications. Leiker et al. emphasized the need for interactivity and coherence in AI-generated materials, particularly in synthetic video production. Pellas (2025), working in the context of science teacher education, noted that problem-based learning benefits from AI-generated videos only when they are contextually adapted to the learners' needs. Arkün-Kocadere and Özhan (2024) highlighted a critical drawback: although students achieved similar academic performance, AI-generated instructors led to lower video engagement, especially in longer content formats.

Zhang et al. (2025) investigates combinations of AI voice, AI avatars, human voice and human avatars in terms of engagement, cognitive load, and perceived usefulness. They show that while both AI voice and AI avatar can improve user engagement independently, a significant improvement in engagement is only observed when both voice and avatar are AI-generated. Zhang and Wu (2024) examine how avatar presence (virtual avatars) in educational videos influences user experience, engagement, and emotional response. Their results show that a realistic and quality avatar can be created by considering several key factors. Dinçer (2022) investigates whether use of synthetic/computer generated voice instead of natural human voice affects comprehension, cognitive load, and learner satisfaction. They demonstrate that high-quality synthesized voices are as effective as human voices for learning outcomes, knowledge retention, and cognitive load. These studies are really encouraging for AI-powered video lecture making.

More practically oriented contributions, such as Zheng (2025) and the HeyGen guide, showcase workflows to convert conventional video content into AI-enhanced formats. Zheng demonstrated how segmenting long-form content into short, engaging clips using generative AI could maintain learner attention. However, such methods often rely on predefined templates and lack deep integration with custom instructional design.

Despite their promise, fully automated video-making platforms, such as, Synthesia, HeyGen, Elai.io, InVideo, Pictory AI, Vyond, Tome AI, AutoLV (Wang, W. 2022), face several limitations. These include lack of pedagogical nuance, rigid script generation, difficulty aligning voice tone with slide visuals, and insufficient flexibility for editing or adapting content to specific teaching contexts. Furthermore, the emotional flattening of synthetic speech and generic presentation styles often lead to disengagement, especially when content complexity demands more human-like expression or emphasis.

To be specific, common auto video-making tools are designed for converting text to video, however, for academic ppt slides full of visual content and animations, those tools share several key weaknesses:

**(a) Lack of Script–Slide Synchronization**

Most tools cannot intelligently align narration with visual slide elements or allow for voice-controlled pacing across topics.

**(b) Poor Visual–Pedagogical Integration**

Automatic tools often prioritize aesthetics over clarity. Educational videos need accurate, step-wise explanations—not just pretty visuals.

**(c) Minimal Instructor Control**

Many platforms restrict editing after video generation. Once a script is processed into video, adjusting voice tone, pacing, or slide coordination can be cumbersome or impossible.

**(d) Limited Interactivity for Students**

Most videos are rendered linearly, giving students no control over navigation by topic or slide, reducing reusability as study materials.

**(e) Emotionally Flat or Robotic Voices**

While improving, voice synthesis often lacks the expressiveness required for conveying subtle distinctions, enthusiasm, or emphasis that human instructors naturally provide.

To address these challenges, we propose a semi-automated approach that combines the strengths of AI with human oversight and pedagogical insight. In this workflow, Google Gemini generates customizable voice scripts, which are then synthesized via Amazon Polly into natural-sounding speech. The resulting voice tracks are synchronized with PowerPoint slides to produce either a timed slideshow or full video lecture. This setup offers unique flexibility: instructors can fine-tune the scripts to match their teaching style, add visual prompts or emphasis where needed, and adapt tone and detail level based on audience. Students benefit from the ability to navigate slide shows at their own pace, while educators retain control over instructional intent, clarity, and accessibility.

In essence, our method bridges the gap between automation and instructional effectiveness. It allows for high-quality video lectures with rich customization—especially vital when aligning audio with complex visual content or when tailoring content for different learning scenarios. By combining AI's efficiency with human-centered design, this hybrid strategy elevates the potential of AI-generated educational content.

The rest of the paper is organized as follows. In Section 2, we formally present our methodology; we then present the experimental results in Section 3. The paper is concluded in Section 4.

# 2. Methodology

The process of our methodology consists of three main components:

- **Script Generation:** Using Google Gemini, instructors prompt the model with high-level lecture topics, slide outlines, or specific concepts. The model outputs structured, editable scripts tailored to the slide content and learning objectives.

- **Voice Synthesis:** The generated scripts are converted to human-like audio using Amazon Polly. Polly's multiple voice options and language support allow fine-tuning of vocal tone, pace, and inflection to suit pedagogical preferences.

- **Video Assembly:** Synthesized voiceovers are synchronized with PowerPoint slides. Videos are exported either as narrated slideshow recordings or compiled screen recordings for full video playback. Optional instructor editing allows for pauses, visual annotations, and voice refinements.

## 2.1 Script Generation Using Google Gemini

Although many generative models and tools are available to generate voice script for academic ppt presentations, Google's Gemini is the most powerful if not the only model to generate both meaningful and accurate script for academic ppt slides with rich visual content. In the following, we demonstrate a few case studies of generated script for ppt slides with complex visual content.

The following prompt is used to generate scripts for the case studies:

*Generate in-depth voice script for the attached ppt slides.*

### *Case 1: Examples of Utilitarianism*

Figure 1 is a ppt slide with images from film posters and a company website, it is one of the 15 slides on the introduction of utilitarianism, which is a lecture section on ethic theories in an IT Professionalism course. The quoted script generated by Gemini is shown in the following shaded text area. As can be seen, with just a single line of tip for the interpretation of the images: Example of Utilitarianism, the script accurately captures the meaning of the images, and the interpretations are well aligned with the definition of utilitarianism.

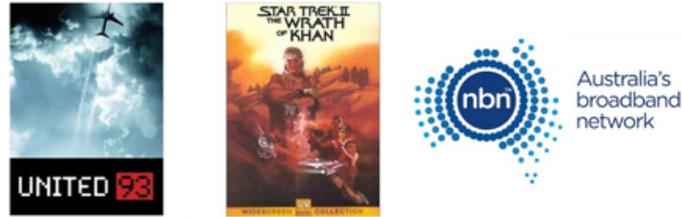

## Examples of Utilitarianism

### What are some examples of using utilitarianism (consequence-based) ethics?

Figure 1. A ppt slide with examples of utilitarianism.

*Script for the ppt slide shown in Figure 1:*

"To help us understand utilitarianism, let's consider some real-world and fictional examples. Looking at the slide, we see images that can spark discussion.

On the left, we have the movie poster for 'United 93,' a film depicting the events of 9/11 where passengers famously fought back against hijackers. This scenario raises complex utilitarian questions: was the decision to resist, potentially sacrificing the lives on board, justified by preventing a greater loss of life on the ground? This is a classic example of evaluating actions based on their consequences for the broader population.

In the middle, we have 'Star Trek II: The Wrath of Khan,' a science fiction classic. In its climax, Spock sacrifices himself to save the ship and crew, uttering the iconic line, 'The needs of the many outweigh the needs of the few, or the one.' This is a quintessential illustration of a utilitarian decision in a fictional context.

On the right, we see the 'nbn' logo, representing 'Australia's broadband network.' Large-scale infrastructure projects like this are often justified on utilitarian grounds. The immense investment and potential disruption are deemed acceptable because the widespread access to high-speed internet is expected to generate significant economic benefits, educational opportunities, and overall societal happiness for a large number of citizens."

*Case 2: Pleasure and Pain*

Figure 2 is a ppt slide with images of a baby in smile and crying, it is also one of the 15 slides on the introduction of utilitarianism, which is a lecture section on ethic theories in an IT Professionalism course. The quoted script generated by Gemini is shown in the shaded text area following the figure. Again, with just a single line of tip for the interpretation of the images: Pleasure and Pain, the script accurately captures the meaning of the images "This slide visually represents this fundamental duality", and the interpretations are well aligned with the extended definition of utilitarianism or Mill's utilitarianism.

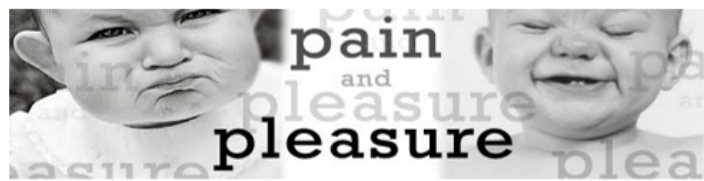

**Pleasure and Pain**

List all the things that bring you
pleasure/happiness and pain. Discuss whether all
the pleasure are good and all the pains are bad.

Figure 2. A ppt slide with examples of pleasure and pain.

*Script for the ppt slide shown in Figure* 2:

"At the heart of utilitarianism is the concept of 'pleasure and pain,' or 'happiness and suffering.' This slide visually represents this fundamental duality.

Utilitarianism posits that happiness is good and pain is bad. The goal is to maximize happiness and minimize pain. However, this raises a profound question: 'Are all pleasures good, and all pains bad?'

This isn't as simple as it seems. For example, the pleasure derived from consuming unhealthy food might lead to long-term pain or health issues. Conversely, the pain of intense exercise or studying for an exam can lead to greater long-term pleasure and well-being. So, it's not just about immediate sensations, but often about the overall, long-term consequences and the quality of the pleasure or pain, which Mill particularly emphasized."

***Case 3: Elements of E-Commerce: Financial Networks***

Figure 3 is a ppt slide with an image of financial transaction network, it is one of the 12 slides on E-commerce elements in a lecture section of an E-commerce Security course. The quoted script generated by Gemini is shown in the shaded text area following the figure. As can be seen, with just a single line of tip for the interpretation of the images: Every link is a security risk, the script accurately captures the meaning of the images, and the interpretations are well aligned with the topic.

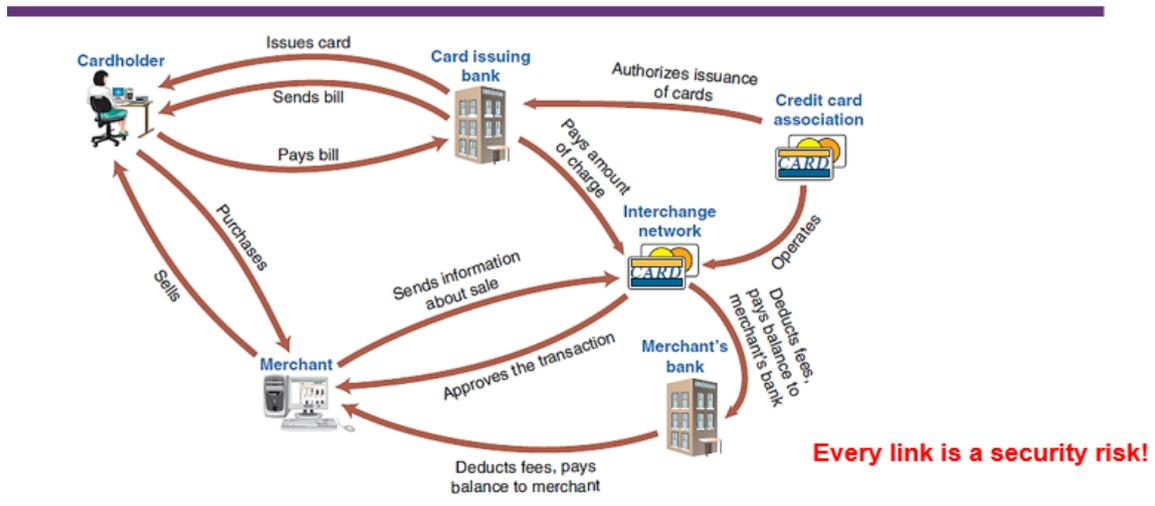

Figure 3. A ppt slide with networks of a financial transaction (Schneider, G (2016), E-Commerce, Course Technology).

*Script for the ppt slide shown in Figure* 3:

"This diagram illustrates a typical financial network for credit card transactions. You can see the flow of information and money between the cardholder, the merchant, the merchant's bank, the card-issuing bank, and the credit card association via the interchange network. It's a complex web, and critically, every single link in this chain represents a potential security risk. This highlights the immense importance of robust security measures in e-commerce financial systems."

***Case 4: Coherent Inference***

Figure 4 shows two linked slides on a case study of applying utilitarianism. The first slide shows the case as an image, and the second slide asks a follow up question. The generated script after Figure 4 successfully connects the case with the discussions in all the previous slides and accurately apply them to the case.

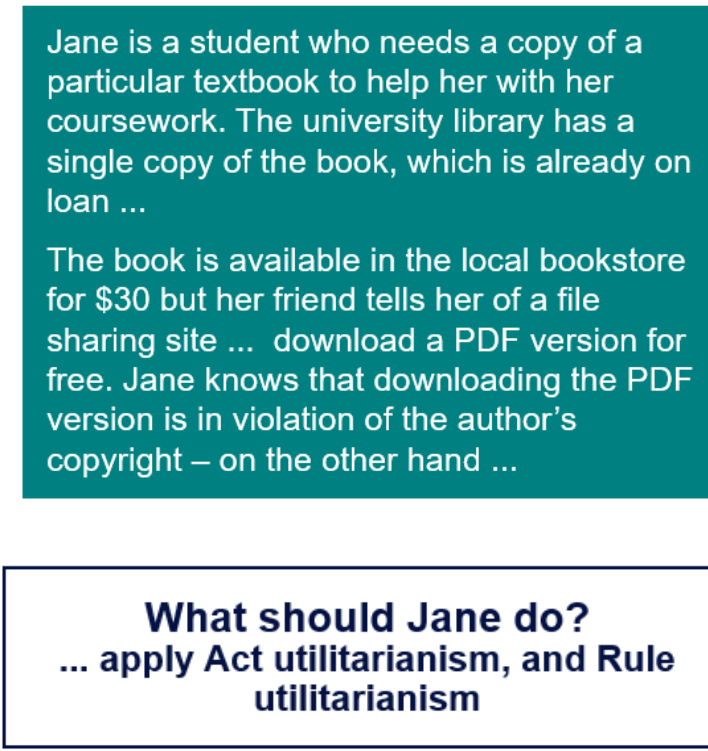

Figure 4. Two linked ppt slides with a case and a follow up question.

*Script for the second ppt slide shown in Figure* 4:

"So, the question we pose to Jane, and to ourselves, is: 'What should Jane do in this situation?'

To analyze this, we need to apply both Act Utilitarianism and Rule Utilitarianism.

From an Act Utilitarian perspective, Jane would weigh the immediate consequences of her individual act of downloading the book. The pleasure or benefit she gains (free access, better grades, reduced financial stress) versus the minimal pain or harm caused to the author or publisher by her single, uncompensated download. A strict act utilitarian might conclude that her individual benefit outweighs the negligible harm, making the download morally permissible in this specific instance.

From a Rule Utilitarian perspective, Jane would ask: 'What if everyone followed the rule: "It's okay to download copyrighted material for free if it benefits you and causes minimal harm to the author"?' If this rule were universally followed, it would lead to widespread copyright infringement, undermine the creative industries, disincentivize authors from producing work, and ultimately cause significant economic damage and reduced societal happiness. Therefore, a rule utilitarian would likely conclude that downloading the PDF is unethical, because the rule it embodies would lead to negative overall consequences for society.

This dilemma starkly illustrates the difference between judging an individual action versus judging the universal rule that action embodies, and it's a common ethical challenge in the digital age."

From the four case studies above, it can be observed that academic presentations usually have high demand for sophisticated visual interpretation. Conventional auto video making tools are typically good at text-to-visualisation, however, they are not designed for visual-to-text interpretation. Google's Gemini is uniquely positioned for scripting academic presentations with rich and visual content, and is the most powerful multi-modal model because it makes use of cross-modal interactions to analyse multimedia content. Other top LLMs including ChatGPT and DeepSeek still lack the capability of doing multi-modal analysis.

With powerful LLM models like Gemini, instructors can also generate scripts with various levels of details to adapt to the nature of the topics or individual slides. To demonstrate the difference, three levels of scripts are generated for the ppt slide on spoofing shown in Figure 5 using the following prompt:

*Generate three versions of voice script for the attached ppt slide with various levels of details: level 1 is normal; level 2 is elaborated; and level 3 is in-depth*

This functionality of generating various levels of details is very useful for instructors to do micro tuning on lecture topics and slides depending on the importance or do compromise between details and time duration.

Figure 5. A ppt slide on spoofing.

*Level* 1 *script*: *normal details*

"Spoofing is a form of cyberattack where a perpetrator pretends to be someone else. This often involves tricking users by substituting a fake website address for a legitimate one using vulnerabilities in Domain Name Servers, or DNS. These servers connect domain names to IP addresses. When compromised, users can be redirected to malicious sites that closely resemble the real ones."

*Level* 2 *script*: *elaborated details*

"Spoofing refers to an attack where someone impersonates a trusted entity to deceive users. One common target is the Domain Name System, or DNS, which acts like the internet's phone book—linking human-friendly domain names to machine-readable IP addresses. Cybercriminals exploit weaknesses in this system to replace the real website's IP address with their own. As a result, when a user types in a trusted URL, they are redirected to a fraudulent site. This fake website may look identical to the original, fooling users into entering personal information or login credentials. This technique is also known as masquerading, as the attacker pretends to be a legitimate entity."

*Level 3 script: in-depth details*

"Spoofing is a sophisticated cybersecurity threat that involves deception at the network level. A common vector for spoofing attacks is the Domain Name System, or DNS, which is a foundational part of the internet's infrastructure. DNS servers are responsible for translating human-readable domain names—like 'example.com'—into numerical IP addresses that computers use to locate each other.

In a spoofing attack, malicious actors exploit software vulnerabilities in DNS servers to manipulate this lookup process. They substitute their own server's IP address in place of the legitimate one. This means when a user tries to access a trusted website, they are silently redirected to a counterfeit version designed to mimic the real thing. These fraudulent sites are often nearly indistinguishable in appearance from legitimate ones, making them effective tools for stealing credentials, spreading malware, or harvesting sensitive data.

This method is also referred to as masquerading, where the attacker adopts a false identity online to gain the user's trust. Unlike simple phishing emails, DNS spoofing occurs behind the scenes, making it harder for end users to detect until damage is done. Understanding and defending against spoofing is critical in any cybersecurity strategy."

### *2.2 Voice Synthesis Using Amazon Polly*

Amazon Polly is a powerful text-speech tool which offers free service for a year with up to 1 million tokens. Users need to create an AWS account before using the service. The interface of Amazon Polly is shown in Figure 6. It is simple, user-friendly and very easy to use. It takes just a couple of steps to setup. First, in the 'Engine' section, select one of the methods to generate the voice.

The next is to select the language or the type of a language and a simulated voice which is named after its human speaker. Then you are ready to go by just pasting the script into the Input text area. You may test the generated audio by listening to the voice before downloading it to a local drive. The audio generation is very fast, typically instant or takes just a couple of seconds.

Amazon Polly offers Speech Synthesis Markup Language or SSML feature which is very useful to create controlled and customizable audios. This feature is activated by clicking the SSML button at the top right of the Input text box. Instructors can use SSML to control the pace of the speaker, add pause between slides etc.

For example, an instructor can use the following SSML to control the speed of the speaker to be 85% of his/her usual speaking speed:

      <speak>

      <prosody rate="85%">

            *My script here ……*

      </prosody>

      </speak>

Before generating audio, script needs to be sanitised to remove/replace unintended characters such as dots and ellipses, '/' and '&' also need to be replaced to avoid clashes with SSML. Or simply use a prompt to explicitly exclude them:

*Generate in-depth voice script for the attached ppt slides and avoid using such characters as dots, ellipses, '/' and '&'*

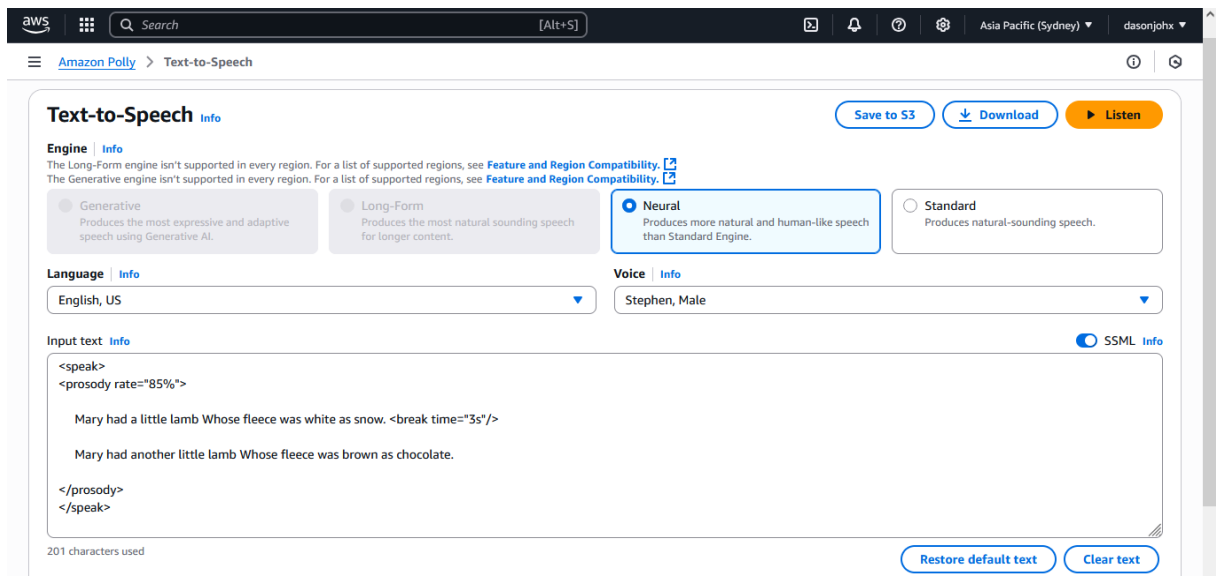

Figure 6. Amazon Polly.

## 2.3 Video Assembly

Long videos are notoriously difficult to make and edit, because mistakes and tech glitches are common during video making, and they increase and accumulate as a video gets longer. It's very costly to correct them during postproduction. Therefore, to make video lectures, a common practice is to break down a two-hour lecture into 4-5 shorter sections with each section having 10-15 ppt slides on average. This approach is much more practical and manageable. Entire video can be made by just using Microsoft PowerPoint.

The first step of making a video lecture is to synthesize the generated audios from Amazon Polly with corresponding ppt slides. This is done by using Insert→Media→Audio as shown in Figure 7 (a), an audio/speaker icon is displayed once the audio is inserted, and it can be moved around.

Click the audio/speaker icon, the Playback setting panel is shown in the ribbon menu (Figure 7 (b)), where an instructor can tell PowerPoint if the audio should start automatically or when clicked on, whether the icon should be hidden during slide show etc.

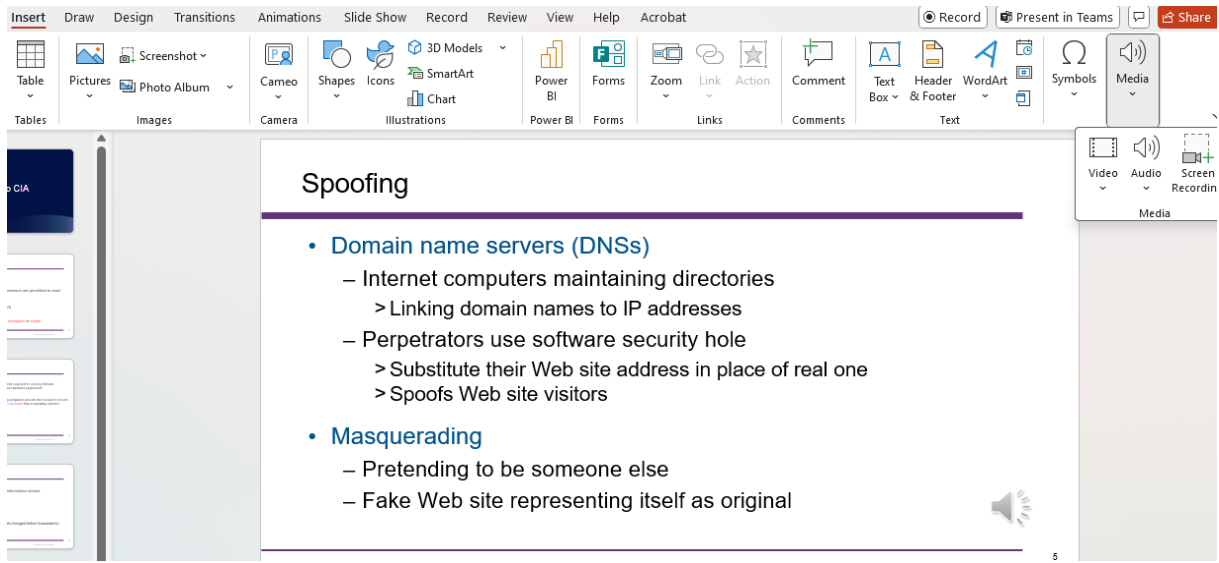

(a) Insert an audio to a ppt slide.

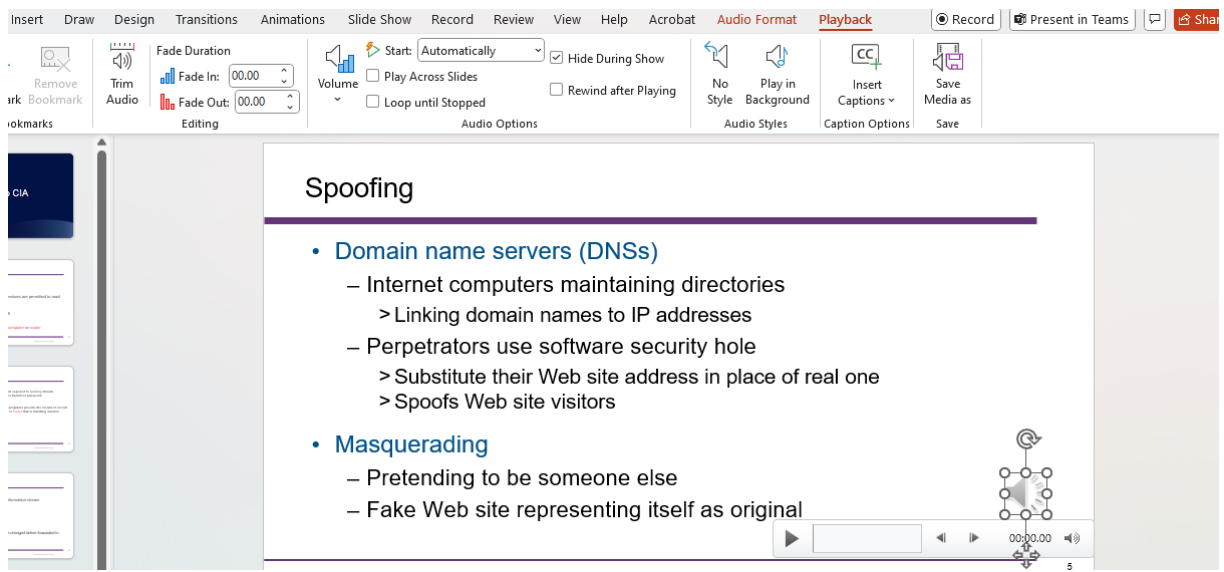

(b) Setting audio playback in a ppt slide.

Figure 7. Synthesize audio into a PowerPoint slide.

Once the voice audios have been synchronized with the corresponding ppt slides and the audio playback has been properly setup, the voiceover PowerPoint is naturally a video in slide show mode. To give students maximum control over the video, the audio/speaker icon needs to be shown at the corner of the slide show, so students can move the mouse over the audio/speaker to stop or replay it. In this case, if a student loses the content-audio synchronization, he/she can simply stop the audio and replay it to catch up.

Script should also be copied into the note area of each ppt slide, so that students can read the script at the right-hand side of the screen when playing the slide show in presenter view as shown in Figure 8. Presenter view can be activated by clicking the three-dot circle at the bottom right of the slide show.

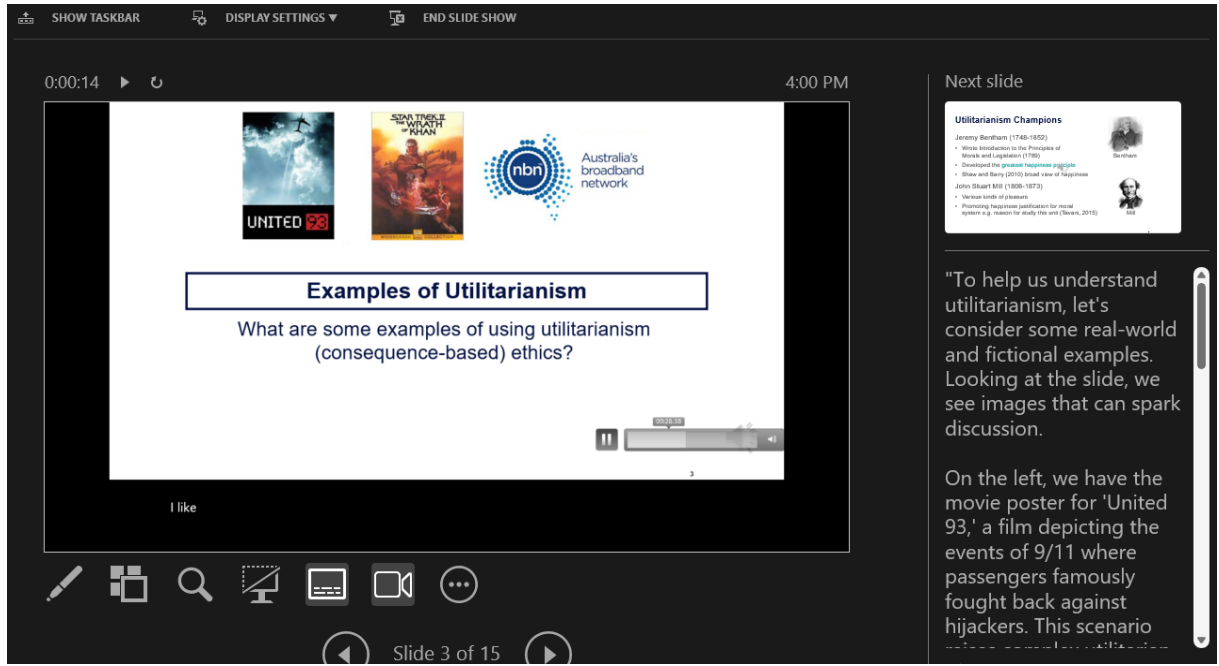

Figure 8. Presenter view of ppt slide show.

The voiceover ppt slides can be converted to a recorded video using the PowerPoint built-in recorder at the top right of the ribbon menu. Click the 'Record' button, the following recording screen is shown (Figure 9). Then, click the red button at the top, the recording is started. At the end of the recording, click the 'Export' button at the top to output the recorded video. One advantage of doing video recording is that an instructor can use the mouse or color tools to draw students' attention.

Instructors can record PowerPoint video with a single script and audio. In this case, the entire script of all the ppt slides is taken as a single input to Amazon Polly and generate a long audio. However, this needs more editing to the script by removing separator information between slides before the audio generation. Furthermore, pause time also needs to be added between slides using the following SSML to allow transition to next slide:

\<speak\>

\<prosody rate="85%"\>

    *Slide* 1 *script* <span style="color:red">*\<break time="3s"/\>*</span>

    *Slide* 2 *script* <span style="color:red">*\<break time="3s"/\>*</span>

    ………

    *Slide* 10 *script*

\</prosody\>

\</speak\>

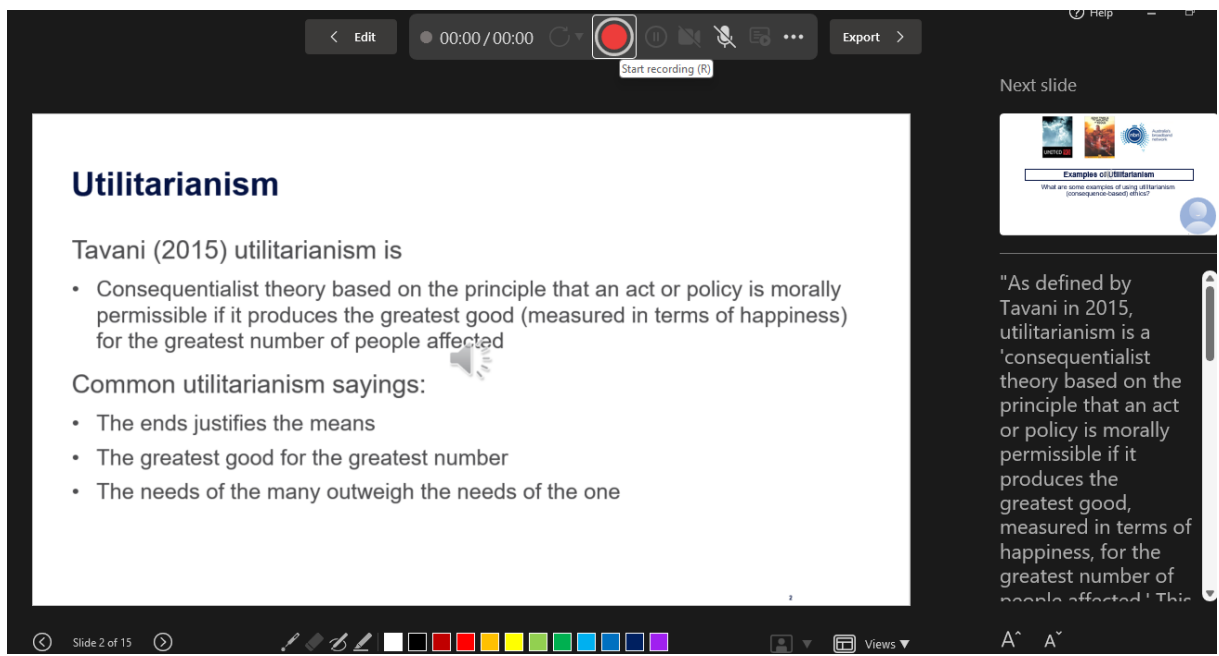

Figure 9. Video recording using PowerPoint built-in recorder.

Provided that the same ppt slides are used, the time taken to generate the script and synthesize audio onto slides is about the same as recording a video by a human instructor. However, for a human instructor to make the video lecture, the time taken to peruse and rehearse (amount to generating script manually and memorising it) the ppt slides before the recording is typically 2-3 times of the actual recording time, plus the time spent on postproduction editing. Therefore making an AI-generated instructional video or **AIIV** is 3-4 times faster than making a human instructional video or **HIV**. This is consistent with other findings in literature [Netland et al 2025][ Leiker et al (2023)].

# 3. Experiments

To compare AIIV with HIV, we have used AIIV on two courses to test student's preference and evaluate its performance against HIV. Both the AIIV and HIV are made using the same ppt slides without avatar or human face, in other words, the only difference between the AIIV and HIV are the voice and narration/script.

## 3.1 Experiments Setup

Two courses are selected for piloting the AIIV for two weeks. The first one is a postgraduate course (Course A) on Leadership in IT Project Management with 37 students and the second one is an undergraduate course (Course B) on IT Professionalism with 134 students. The AIIV videos are made in both slide show and recorded formats as described in Section 2.3. The number of slides for each of the videos ranges from 8-12 and the duration of the videos is between 5-9 minutes.

Both qualitative and quantitative data are collected for the experiments. In the following, we describe the two types of experiments in details.

## 3.2 Qualitative Experiment

For qualitative experiment, a survey is conducted by using the following questionnaire. To collect as many responses as possible, the questionnaire is designed to focus on 8 key aspects of learning and the answer to each of the questions is simplified as Yes or No.

***Questionnaire***:

1. Conceptual Clarity: Y/N

   • Does the video explain concepts in a way that is understandable and logically sequenced?

2. Cognitive Load Balance: Y/N

   • Is the information density manageable without overwhelming the learner?

3. Use of Examples and Analogies: Y/N

- Does the video effectively use examples, visuals, or analogies to clarify abstract ideas?

4. Alignment with Learning Objectives: Y/N

    - Are the video contents clearly aligned with the topic outcomes?

5. Narrative Coherence: Y/N

    - Is the script logically structured, with clear transitions between topics or ideas?

6. Ease of Use: Y/N

    - Was it easy to access, play, and navigate the video content?

7. Preference Over Alternatives: Y/N

    - Do you prefer this video format over live lectures or human instructional videos?

8. Preference of PowerPoint Slide Show over PowerPoint Recording: Y/N
    - Do you prefer the slide show with voiceover over the recorded PowerPoint video?

Altogether, 56 responses from the survey are received. The following table shows the statistics of the Yes responses:

Table 1. Percentage of Yes responses to the questionnaire

| Question | 1 | 2 | 3 | 4 | 5 | 6 | 7 | 8 |
|---|---|---|---|---|---|---|---|---|
| Yes | 100% | 96.4% | 100% | 100% | 98.2% | 96.4% | 91% | 92.8% |

The following are the comments (sic) received from the participants. As the comments (or reasons) are optional, not all participants leave comments.

1. Conceptual Clarity: Y/N

    - I think the concepts in the new video lectures are explained more clearly and concisely, reducing unnecessary information and helping to grasp the key points.

    - The structure of the content makes sense, explaining foundational concepts and building upon these (as did the previous video recordings).

    - The Content is better paced sequentially.

    - The concepts were explained clearly and flowed in a logical way.

- The movie breaks down ideas in a way that makes sense, which makes them easier to understand and follow.

- The videos broke down tricky concepts into crystal clear, step-by-step explanations that were super easy to follow.

2. Cognitive Load Balance: Y/N

- I think the new video lectures have a reasonable information density and an appropriate pacing, which helps reduce cognitive load while watching. However, I still need to research some technical knowledge before I can understand it.

- The pace of the narrator is the right balance between slow enough to allow for ideas and concepts to sink in. But not too slow that it feels laborious. The pace of the narrator is the right balance between slow enough to allow for ideas and concepts to sink in. But not too slow that it feels laborious.

- The Idea tracking is easier to follow, and when using.

- The amount of information was just right—not too overwhelming.

- There are reasonable chunks of information, so it's easy to take in without feeling too much.

- The info was perfectly balanced, packed with insights but never overwhelming, keeping me hooked throughout.

3. Use of Examples and Analogies: Y/N

- I think the added real-world examples and analogies in the new videos help to make abstract concepts more concrete, making them easier to understand.

- The audio content was good and provided sufficient examples.

- The added examples and analogies do help to clarify the ideas, and make it easier to understand even if not directly listening.

- Examples and visuals helped make the ideas easier to understand.

- The similarities and examples help to make abstract ideas clearer, which makes difficult topics easier to understand.

- Awesome examples and visuals made tough ideas feel simple and stick in my mind like glue.

4. Alignment with Learning Objectives: Y/N

- I think the video content accurately aligns with the learning objectives of each lecture and does not stray from the course topic.

- The Contents are clearly aligned, and the divergents from the slides are relevant to the Outcomes.

- The content matched what we were meant to learn for each topic.

- The course goals are directly linked to the content, so the video helps students learn what they need to learn.

- Every part of the videos hit the topic's goals spot-on, making learning feel focused and meaningful.

5. Narrative Coherence: Y/N

- The narrative flow is smooth and natural, with each segment connecting closely together. But I think the pace is still a bit fast.

- The structure of the content makes sense, explaining foundational concepts and building upon these

- The Sound quality is better and clearer, with easier-to-follow tracking of ideas.

- The structure made sense and transitions between topics were smooth.

- The script is organized in a way that makes sense, and the video flows easily because of this.

- The script flowed like a story, with smooth transitions that tied everything together beautifully.

6. Ease of Use: Y/N

- I think the new video interface has a clear layout, intuitive navigation, and is easier to use, which improves the overall viewing experience.

- I say NO with mixed feelings. On a PC with PowerPoint, viewing the content and accessing the presenter notes was easy. However, I went with NO as when trying to view on a mobile device, the notes were not readily accessible, and the audio was playing twice with a delay on the second track, making it incomprehensible.

- For the most part, it followed the Slide content. When it did diverge, I would have to take a moment to identify if it's adlibbing the next line or presenting a new idea. This is the same with the Human Instructional videos. It is easier to get transcripts for, however.

- It was easy to access and navigate through the video without any issues.

- The layout makes it easy to get to the movie and move around in it, and watching it is smooth.

- Playing and navigating the videos was a breeze, with no hiccups, making learning pure joy.

7. Preference Over Alternatives: Y/N

- I think the new videos are more visually appealing and interactive than traditional videos. This method helps me learn more quickly, allowing me to follow along step by step.

- I only say NO for the fact of the AI-generated voice in place of a human. Otherwise, I really like the PowerPoint with the transcript and pausing after each slide. I think it is moving in a good direction. But would rather a more human experience from the lectures. And value being able to see the speaker's face. I miss face-to-face classes and felt these presentations lacked that human connection.

- The Format is easier to understand and comprehend in comparison.

- It felt a bit robotic and plain—I'd still prefer a human presenter, but the voice in this format has more clarity, which makes it work well.

- This format is better than live lectures or other instructional movies because it is easier to understand and follow.

- These videos were way more fun and clear than live lectures or other formats—total game-changers!

- Compared with the original video, the sound is clearer, and it would be better if the speech speed is slightly slowed down.

8. Preference of PowerPoint Slide Show over PowerPoint Recording: Y/N
- I think recordings, with their audio narration and dynamic emphasis, are easier to understand and remember than just showing slides.

- Having access to the full transcript in the PowerPoint slides makes it easier to copy names or terms.
- The Slideshow allows for easier note-taking.
- I prefer the PowerPoint recording, but only if the person explains the slides instead of just reading them.
- The slideshow with speech makes it easier to take notes and understand because it makes the most important points stand out.
- The slide show with voiceover was lively and engaging, making it way more exciting than a plain recording.

Overall, the qualitative experiment shows that the responses from students are highly positive about the AIIV. This is encouraging as the new type of video making are much more efficient than traditional HIV, manifesting a large scale of application of the new AI-powered technology across high education.

However, based on students' feedback, there is significant room for improvement in both audio quality and the addition of avatars to the videos. At this moment, it's an initiative of proof of concept by using free tech service. In the next stage, more natural and high-quality audio generating tools will be explored, natural avatars or instructor's talking head can also be generated to synchronize with the audio. By improving these two aspects with minimal funding, it's foreseeable that AIIV will replace HIV to become a prevalent practice in high education in the next few years.

### 3.3 Quantitative Experiment

To conduct the quantitative experiment, we collected results from the first 2 quizzes of each of the two courses and compared them with the results from the previous semester which was taught by the same lecturer using the same materials except the video lectures, whereas the HIV videos were used in the first semester while the AIIV videos are used in the first two weeks of the second semester. The HIV videos are recorded by a native English speaker who has taught the courses.

For Course A, the maximal mark for each quiz is 3, while for Course B, the total mark for each quiz is 10. The following tables show the statistics of the two courses over two semesters.

***Course A***.

Table 2. Statistics of two quizzes from each semester

|  | Semester 1 (67 counts) | | Semester 2 (37 counts) | |
|---|---|---|---|---|
| Quiz | 1 | 2 | 1 | 2 |
| Mean ($\mu$) | 2.13 | 2.25 | 2.21 | 2.09 |
| Standard deviation ($\sigma$) | 0.33 | 0.29 | 0.56 | 0.48 |

***Course B***.

Table 3. Statistics of two quizzes from each semester

|  | Semester 1 (35 counts) | | Semester 2 (134 counts) | |
|---|---|---|---|---|
| Quiz | 1 | 2 | 1 | 2 |
| Mean ($\mu$) | 8.56 | 8.72 | 8.46 | 9.02 |
| Standard deviation ($\sigma$) | 2.13 | 1.20 | 1.91 | 1.11 |

***Hypothesis Test***.

To find out if there is any difference between the results of the two semesters, we have conducted a hypothesis test on the two quizzes of the two courses. In the following, we use the statistics of the first quiz of Course A to demonstrate the procedures of the hypothesis test. Since the two samples have unequal variances and unequal sample sizes, the two-sample Welch *t*-test is used to do the hypothesis test.

Data 1: $\mu_1 = 2.13$, $\sigma_1 = 0.33$, $n_1 = 67$

Data 2: $\mu_2 = 2.21$, $\sigma_2 = 0.56$, $n_2 = 37$

*Hypotheses*:

Null hypothesis H$_0$: $\mu_1 = \mu_2$ (means are equal)

Alternative hypothesis H$_a$: $\mu_1 \neq \mu_2$ (means are different)

*Test*:

a. Pooled standard error:

$$SE = \sqrt{\frac{\sigma_1^2}{n_1} + \frac{\sigma_2^2}{n_2}} \approx 0.100504 \tag{1}$$

b. *t*-score:

$$t = \frac{\sigma_2 - \sigma_1}{SE} \approx 0.796 \tag{2}$$

c. Welch degrees of freedom *df*:

$$df = \frac{\left(\frac{\sigma_1^2}{n_1} + \frac{\sigma_2^2}{n_2}\right)^2}{\frac{\left(\frac{\sigma_1^2}{n_1}\right)^2}{n_1 - 1} + \frac{\left(\frac{\sigma_2^2}{n_2}\right)^2}{n_2 - 1}} \approx 50 \tag{3}$$

d. Two-tailed *p*-value (*t* with $df \approx 50$):

$$p \approx 0.43$$

e. Conclusion:

$$t \approx 0.796,\ df \approx 50,\ p \approx 0.43$$

- At $\alpha = 0.05$, $p = 0.43 > 0.05$, so we fail to reject H$_0$.
- There is no significant difference between Data 1 and Data 2 means.

The *p*-value represents the evidence of accepting H$_0$, the larger the *p*-value, the more evidence of accepting H$_0$ which indicates that there is no significant difference between the two populations under test. In practice, it is compared with the $\alpha$-value which represents the confidence level (1- $\alpha$) of the test. If $p > \alpha$, H$_0$ is accepted, otherwise H$_a$ is accepted.

The following table show the results of hypothesis testing of the two quizzes of the two courses. It can be observed that in all the 4 tests, none of them has significant difference, indicating that the AIIV has the comparable or equal performance to the HIV.

Table 4. Results of hypothesis testing on the two courses

|  | Course A | | Course B | |
|---|---|---|---|---|
| Quiz | 1 | 2 | 1 | 2 |
| $p$-value | 0.43 | 0.07 | 0.80 | 0.19 |
| $\alpha$ | 0.05 | 0.05 | 0.05 | 0.05 |
| $H_0$ | ✓ | ✓ | ✓ | ✓ |
| $H_a$ | | | | |

## 4. Conclusions

This study demonstrates that AI-powered video lecture production can significantly improve efficiency while maintaining instructional quality. By integrating script generation, voice synthesis, and slide synchronization into a streamlined workflow, educators can create instructional videos that rival traditional human-recorded content in effectiveness, while being three to four times faster to produce.

The pilot experiments show no significant difference in student performance between AIIV and HIV formats, underscoring the viability of AI-generated videos as mainstream teaching tools. Student feedback further highlights strengths in clarity, pacing, and conceptual alignment, though concerns regarding the robotic nature of voices and the lack of visual presence remain.

These findings point to a promising future where AIIVs can complement or even replace HIVs in many teaching contexts, provided ongoing improvements are made in audio naturalness and avatar integration. With modest investment in more advanced synthetic media tools, AI-powered video lectures have the potential to become a standard practice in higher education, expanding access, reducing production barriers, and supporting diverse learning environments.